\documentstyle[aps,eqsecnum,amssymb,amsbsy,preprint,tighten,epsfig]{revtex}
\title{
{\hfill {\small \begin{tabular}{r} CGWP 01/9-1 \\ CGPG 01/9-1 \end{tabular}} } 
\vspace*{2cm} \\
Bounding the mass of the graviton using binary pulsar observations}
\author{Lee Samuel Finn\thanks{Also Center for 
Gravitational Physics and Geometry, Department 
of Physics, and Department of Astronomy and Astrophysics; e-mail 
lsf@gravity.phys.psu.edu} 
and Patrick J. Sutton\thanks{Also Center for Gravitational Physics and 
Geometry and Department of Physics; e-mail psutton@gravity.phys.psu.edu}
}       
\address{
Center for Gravitational Wave Physics, 
The Pennsylvania State University,
State College, PA, USA 16802-6300.
}

\begin{document}


\maketitle

\begin{abstract}
The close agreement between the predictions of dynamical general
relativity for the radiated power of a compact binary system and the
observed orbital decay of the binary pulsars PSR~B1913+16 and
PSR~B1534+12 allows us to bound the graviton mass to be less than 
$7.6\times10^{-20}\,\text{eV}/c^2$ with 90\% confidence. This bound 
is the first to be obtained from dynamic as opposed to static field 
relativity.  The resulting limit on the graviton mass is within two orders of
magnitude of that from solar system measurements, and can be expected
to improve with further observations.
\end{abstract}
\pacs{04.30.-w, 04.80.-y, 14.70.-e, 97.80.-d}


\section{Introduction}
\label{Introduction}
 
General relativity assumes that gravitational forces are propagated by
a massless graviton.  Current experimental limits on the graviton mass
are based on the behavior of static gravitational fields.  In
particular, a nonzero graviton mass $m$ would cause the gravitational
potential to tend to the Yukawa form $r^{-1}e^{-mr}$, effectively
cutting off gravitational interactions at distances greater than the
Compton wavelength $m^{-1}$ of the graviton.  The absence of these
effects in the solar system \cite{talmadge88a} and in galaxy and
cluster dynamics \cite{goldhaber74a,hare73a} thus provides an upper
limit on $m$.

In the dynamical regime, a nonzero graviton mass would produce 
several interesting effects.  These include extra degrees of freedom 
for gravitational waves (e.g., longitudinal modes), and propagation at 
the frequency-dependent speed 
\begin{equation}
\label{speed}
v = \sqrt{1-m^2/\omega^2} \, .  
\end{equation}
Recently, Will \cite{will98a} and Larson and Hiscock \cite{larson00a} have
proposed techniques for examining the latter effect with future
gravitational-wave interferometer observations to place a limit on
$m$.  Here we present a new method of bounding the graviton mass, 
which makes use of existing binary pulsar observations.  Our
technique is based on the agreement between the observed orbital decay
of the binary pulsars PSR~B1913+16 and PSR~B1534+12 and the predictions of
general relativity \cite{taylor94a,Wo:91,stairs99a}.  This is the first 
bound on $m$ from dynamic-field relativity to be accessible with existing 
observational data, and it provides a limit that is independent of the 
Yukawa bounds.

The idea is quite simple.  Consider the Hulse-Taylor binary pulsar, 
PSR~B1913+16, of which the observed decay rate  
coincides with that expected from relativity to approximately $0.3\%$. 
A nonzero graviton mass would upset this remarkable agreement by
altering the predicted orbital decay.\footnote{
Corrections to other characteristics of the system are 
negligible by comparison on dimensional grounds: 
$(mr)^2=(m/\omega)^2 (v/c)^2$, $(mM)^2=(m/\omega)^2(v/c)^6$, 
where $v/c = {\cal O}(10^{-3})$ for these binary systems.
} 
This implies an upper limit on
the graviton mass.  A crude estimate of this bound is quickly obtained
from dimensional analysis.  For a system with characteristic frequency
$\omega$ one expects the effects of a graviton mass to appear at
second order in $m/\omega$, as in (\ref{speed}).  For gravitational
waves at twice the orbital frequency of PSR~B1913+16, requiring
$(m/\omega)^2 < 0.003$ implies an upper limit of 
order $10^{-20}\,\text{eV}/c^2$. 
This is comparable to the best limit from solar system observations,
$m<0.44 \times 10^{-21}\,\text{eV}/c^2$ \cite{talmadge88a}.  The purpose of
this paper is to refine and make rigorous this estimate.

In section~\ref{framework}
we discuss linearized general relativity with a massive graviton.  The
field equation and the effective stress tensor for the metric
perturbations (gravitational waves) are found.  In
section~\ref{solutions} we solve the field equations using Fourier
techniques, and derive the gravitational-wave luminosity of a general
slowly moving periodic source when the graviton is massive.  We apply
this result to the observed orbital decay of the binary pulsars 
PSR~B1913+16 and PSR~B1534+12 to obtain an upper limit on the mass 
of the graviton in section~\ref{binarypulsars}, and conclude with some 
brief comments in section~\ref{conclusions}.


\section{Linearized General Relativity with a Massive Graviton}
\label{framework}

In linearized general relativity one writes the metric as a perturbation 
of the Minkowski metric: 
\begin{equation}\label{h}
g_{\mu\nu} = \eta_{\mu\nu} + h_{\mu\nu} \, , \qquad |h_{\mu\nu}| \ll 1 \, .
\end{equation}
We adopt the convention that indices of $h_{\mu\nu}$ are raised and 
lowered using the Minkowski metric; e.g., 
\begin{equation}
    h^\mu_{\hphantom{\mu}\nu} \equiv \eta^{\mu\lambda} h_{\lambda\nu}.
\end{equation}
The linearized theory is defined by substituting (\ref{h}) into the 
Einstein action, expanding in powers of $h_{\mu\nu}$, and keeping only 
terms up to second order in $h_{\mu\nu}$ (giving field equations linear 
in $h_{\mu\nu}$).  

We wish to examine an extension of linearized general relativity which
includes a mass term for the graviton.  We choose the unique mass term
for which the wave equation of the linearized theory takes the
standard form with an $h$-independent source, and for which the
predictions of massless general relativity are recovered by setting
$m\to0$ at the end of the calculations (see
\cite{visser98a,boulware72a} and Appendix~\ref{massterms}).  Following
the procedure described above, we arrive at the action
\begin{mathletters}
\begin{eqnarray}
\lefteqn{I~=~\frac{1}{64\pi} \int \!d^4x \left[ \, \vphantom{\frac{1}{1}} 
             h_{\mu\nu,\lambda} h^{\mu\nu,\lambda} 
             -2 h_{\mu\nu}^{\hphantom{\mu\nu},\nu}
                 h^{\mu\lambda}_{\hphantom{\mu\lambda},\lambda} 
             +2 h_{\mu\nu}^{\hphantom{\mu\nu},\nu}
                 h^{,\mu} 
             \right. }\nonumber \\
  &   &      \hspace{0.0in}\left. \mbox{}       
             -h^{,\mu} h_{,\mu} 
             -32\pi h_{\mu\nu} T^{\mu\nu}  
	     +m^2(
                 h_{\mu\nu}h^{\mu\nu}-\frac{1}{2}\,h^2
	     )
         \, \right] \, , \label{action}
\end{eqnarray}
where 
\begin{equation}
    h \equiv h^\nu_{\hphantom{\nu}\nu}.
\end{equation}
\end{mathletters}
The first five terms are the linearized Einstein action and the stress
tensor source for the metric perturbations, while the last term is our
(phenomenological) choice of mass term \cite{visser98a}.  Linearized
general relativity is regained by setting $m=0$.  At linear order the
stress tensor is assumed to be independent of $h_{\mu\nu}$ and
conserved:
\begin{equation}\label{conservation}
T_{\mu\nu}^{\hphantom{\mu\nu},\nu} = 0 \, .
\end{equation}

The field equations arise from requiring the action to be invariant  
under variations of the metric perturbation; one finds  
\begin{eqnarray}
&&\Box h_{\mu\nu} 
- h_{\mu\hphantom{\lambda},\lambda\nu}^{\hphantom{\mu}\lambda}  
- h_{\nu\hphantom{\lambda},\lambda\mu}^{\hphantom{\nu}\lambda} 
+ h_{,\mu\nu} 
+ \eta_{\mu\nu} 
h^{\rho\sigma}_{\hphantom{\rho\sigma},\rho\sigma}\nonumber\\
&&\qquad{}- \eta_{\mu\nu} \Box h
- m^2(h_{\mu\nu}-\frac{1}{2}\eta_{\mu\nu} h)
  \, = \, -16\pi T_{\mu\nu}  \, . \label{linFE}
\end{eqnarray}
This rather cumbersome equation simplifies considerably when expressed 
in terms of the trace-reversed metric 
perturbations $\bar{h}_{\mu\nu}$, defined by 
\begin{equation}\label{htohbar}
\bar{h}_{\mu\nu} 
  =  h_{\mu\nu} - \frac{1}{2} \eta_{\mu\nu}h \, .
\end{equation}
The conservation of the stress tensor 
requires the divergence of both sides of (\ref{linFE}) to vanish.  
This implies that the mass term itself must have vanishing divergence:
\begin{equation}\label{Lorentzgauge}
\bar{h}^{\mu\nu}{}_{,\nu} = 0  \, . 
\end{equation}
This is equivalent to the Lorentz condition of the massless theory. 
Here, however, it is not a gauge choice; rather, it represents the constraints 
provided by the equations of motion and thus eliminates four of the
ten independent $h_{\mu\nu}$.  The remaining six components represent
true degrees of freedom in the massive theory, which consist of the
five helicity states of the spin-2 field, plus an additional spin-0
component \cite{boulware72a}.

Imposing (\ref{Lorentzgauge}), the field equation may be simplified to 
\begin{equation}\label{fieldeqn}
(\Box-m^2) \bar{h}_{\mu\nu} =  -16\pi T_{\mu\nu}  \, ,
\end{equation}
which is the familiar form of the wave equation for a massive field. 
This will be very convenient for calculations of gravitational
radiation in the massive-graviton theory.  As described above, our
mass term is the unique choice for which the wave equation takes this
standard form with an $h$-independent source, and for which the
predictions of massless general relativity are recovered by setting
$m\to0$ at the end of the calculations (see
\cite{visser98a,boulware72a} and Appendix~\ref{massterms}).
The second property is nontrivial since, for example, the pure spin-2 theory 
with five degrees of freedom does not satisfy the classical 
weak-field tests of general relativity as $m\to0$ 
\cite{boulware72a,VaVe:70,Za:70}.

To analyze the energy content of gravitational waves we need an 
effective stress tensor for metric perturbations.  Applying Noether's 
theorem \cite{wald84a} to the Lagrangian of (\ref{action}) we find  
\begin{eqnarray}\label{canonical1}
T^{\text{GW}}_{\mu\nu}
  & = &  \langle\frac{\delta {\cal L}}{\delta(h^{\alpha\beta,\mu})} \, 
         h_{\alpha\beta,\nu} - \eta_{\mu\nu} {\cal L} \rangle
	 \nonumber \\
  & = &  \frac{1}{32\pi}\langle 
             \bar{h}_{\alpha\beta,\mu} \bar{h}^{\alpha\beta}{}_{,\nu}
             -\frac{1}{2}\bar{h}_{,\mu}\bar{h}_{,\nu}
         \rangle \, . 
\end{eqnarray}
Here the brackets denote an averaging over at least one period of the 
gravitational wave.  Equation (\ref{canonical1}) is identical in form 
to the usual effective stress tensor for gravitational waves with 
$m=0$ \cite{misner73a}.


\section{Solutions}
\label{solutions}

In linearized general relativity the field equation 
(\ref{fieldeqn}) with $m=0$ has the general solution \cite{misner73a} 
\begin{eqnarray}
\bar{h}_{\mu\nu}(t,\vec{x}) 
  & = &  4\int \! d^3\!x'\, 
         \frac{T_{\mu\nu}(t-|\vec{x}-\vec{x}'|,\vec{x}')}{|\vec{x}-\vec{x}'|} 
	 \, . \label{exactsoln} 
\end{eqnarray}
For a massive graviton (\ref{exactsoln}) is no longer applicable, since 
the speed of propagation of the gravitational waves is 
frequency-dependent and so the retarded time 
\mbox{$t-|\vec{x}-\vec{x}'|/v(\omega)$} 
is different for each frequency component of the wave.   
We evade this difficulty by solving (\ref{fieldeqn}) in frequency space, 
dealing with each frequency separately.  A similar analysis 
encompassing the radiation of general scalar and vector fields 
can be found in \cite{krauss94a}. 

In the frequency domain, the field equation (\ref{fieldeqn}) becomes 
\begin{equation}\label{FTfieldeqn}
\left(\nabla^2+[\omega^2-m^2]\right)
\widetilde{\bar{h}}_{\mu\nu}(\omega | \vec{x}) 
  =  -16 \pi \widetilde{T}_{\mu\nu}(\omega | \vec{x}) \, ,
\end{equation}
where the tilde denotes the Fourier transform and $\nabla^2$ is the 
3-space Laplacian.  Equation (\ref{FTfieldeqn})   
is the inhomogeneous Helmholtz equation; the retarded 
Green function $\widetilde{G}_R$ for this equation is
\begin{equation}\label{FTG}
\widetilde{G}_R(\omega|\vec{x};\vec{x}') 
  =  \frac{e^{ik|\vec{x}-\vec{x}'|}}{4\pi |\vec{x}-\vec{x}'|}  \, ,
\end{equation}
where 
\begin{equation}
k\equiv \mbox{sign}(\omega)\sqrt{\omega^2-m^2}
\end{equation}
for $|\omega| > m$.
(The wavenumber $k$ should not be confused with a spatial index.)  
The retarded solution of (\ref{FTfieldeqn}) for fixed $\omega$ is then 
\begin{equation}\label{FTgensoln}
\widetilde{\bar{h}}_{\mu\nu}(\omega | \vec{x}) 
  =  16\pi \int\!d^3\!x' \, \widetilde{G}_R(\omega|\vec{x};\vec{x}') \,
     \widetilde{T}_{\mu\nu}(\omega | \vec{x}') \, . 
\end{equation}

In order to evaluate (\ref{FTgensoln}) we make use of the 
slow-motion approximation, $\omega a \ll 1$, with $a$ the 
characteristic size of the source.  With this assumption, 
and taking the observation point far from the source 
region ($r\equiv|\vec{x}| \gg |\vec{x}'|$), the Green function  
$\widetilde{G}_R$ may be expanded for large $r$.  One finds 
\begin{eqnarray}
\lefteqn{\widetilde{\bar{h}}_{\mu\nu}(\omega | \vec{x}) 
    =    \frac{4e^{ikr}}{r} \int\!d^3\!x' \, 
         \widetilde{T}_{\mu\nu}(\omega | \vec{x}') 
         \left[ \, \vphantom{\left(\frac{\vec{x}\cdot\vec{x}'}{r}\right)^2}
             1 + (-ik)\frac{\vec{x}\cdot\vec{x}'}{r}  
             \right. }\nonumber \\
  &   &      \left. \mbox{}
	     + \frac{1}{2} (-i k)^2 
	     \left(\frac{\vec{x}\cdot\vec{x}'}{r}\right)^2 
         \, \right]   
         \, \left[ \, 
	     1 + {\cal O}\left(\frac{a}{r},(\omega a)^3\right) 
	 \, \right]
	 \, . \label{FTintegral} 
\end{eqnarray}
In the $m=0$ case one can write the metric perturbations due to 
a slowly moving source in terms of the mass $M$, dipole moment $D_j$, 
and quadrupole moment $I_{jk}$ of the source, where 
\begin{mathletters}
\begin{eqnarray}
    M  &=&  \int \! d^3\!x\, T_{00}            \, , \label{mass}   \\
   D_j &=&  \int \! d^3\!x\, T_{00}\,x^j       \, , \\
I_{jk} &=&  \int \! d^3\!x\, T_{00}\,x^j\,x^k  \, . \label{quadrupole}
\end{eqnarray}
\end{mathletters}
We can obtain an analogous result in the frequency domain, 
using the conservation of the stress tensor 
to write the integral over $\widetilde{T}_{\mu\nu}$ in (\ref{FTintegral}) 
in terms of the multipole moments of the source.  
In the frequency domain the conservation equation (\ref{conservation}) 
for the stress tensor becomes 
\begin{equation}\label{FTconserv1}
-i\omega \widetilde{T}_{00} = \partial_j \widetilde{T}_{0j} \, ,  \qquad 
-i\omega \widetilde{T}_{0i} = \partial_j \widetilde{T}_{ij} \, . 
\end{equation}
Using these relations and the slow-motion
approximation, one can show that  
\begin{eqnarray}
\widetilde{\bar{h}}_{00}(\omega | \vec{x})
  & = &  \frac{4e^{ikr}}{r} \left[ \, 
             \widetilde{M} 
             +\frac{x^j}{r} (-ik) \widetilde{D}_{j} 
	     +\frac{x^jx^k}{2r^2}(-ik)^2 \widetilde{I}_{jk}
	 \, \right]  , \nonumber \\
\widetilde{\bar{h}}_{0j}(\omega | \vec{x})
  & = &  \frac{4e^{ikr}}{r} \left[ \, 
             -(-i\omega) \widetilde{D}_{j} 
	     -\frac{x^k}{2r}(-ik)(-iw) \widetilde{I}_{jk}
	 \, \right]  , \nonumber \\
\widetilde{\bar{h}}_{jk}(\omega | \vec{x}) 
  & = &  \frac{4e^{ikr}}{r} \left[ \, 
             \frac{1}{2} (-i\omega)^2 \widetilde{I}_{jk}  
	 \, \right]  , \label{soln} 
\end{eqnarray}
where $\widetilde{M}$, $\widetilde{D}_j$, $\widetilde{I}_{jk}$, are 
respectively the Fourier transforms or Fourier coefficients
of the mass, dipole moment, and quadrupole moment of the source.  
Only the quadrupole terms are relevant to us; the 
mass and dipole moments are constant to linear order in $h$  
[the energy and momentum carried away by the radiation field 
are ${\cal O}(h^2)$], 
hence $\widetilde{M}$ and $\widetilde{D}_j$ contain only zero-frequency 
components and will not contribute to the radiation. 

The rate of energy loss by the source can be found by integrating 
the outward gravitational-wave flux over a sphere centered on 
the source:  
\begin{eqnarray}
L \equiv -\frac{d E}{d t} 
  & = &  \int d\Omega\,r^2 \,T^{0i}_{\text{GW}} \frac{x^i}{r} \, .
\end{eqnarray}
Let us assume the source is periodic with period $P$.  Then the metric 
perturbations $\bar{h}_{\mu\nu}(t,\vec{x})$ in the time domain are 
related to their Fourier 
components $\widetilde{\bar{h}}_{\mu\nu}(\omega,\vec{x})$ via 
\begin{eqnarray}
\bar{h}_{\mu\nu}(t,\vec{x}) 
  &=&  \sum_{n=-\infty}^\infty \widetilde{\bar{h}}_{\mu\nu}(\omega_n,\vec{x}) 
     \, e^{-i\omega_n t} \, , 
\label{sum} \\
\widetilde{\bar{h}}_{\mu\nu}(\omega_n,\vec{x}) 
  &=&  \frac{1}{P}\,\int_{0}^P \!\!dt \, \bar{h}_{\mu\nu}(t,\vec{x}) 
     \, e^{i\omega_n t} \, , 
\label{invsum}
\end{eqnarray}
where 
\begin{equation}
\omega_n = n \, \frac{2\pi}{P} \, ,
\end{equation}
and the tilde now represents a Fourier {\em coefficient}.  Substituting
(\ref{soln}) and (\ref{sum}) into the expression (\ref{canonical1})
for the stress tensor of the gravitational waves the luminosity is
found to be
\begin{mathletters}\label{eqn:Eloss} 
\begin{eqnarray}
    L & = & L_{\text{GR}} + \sum_{n=1}^\infty 
    \frac{m^2 \omega_n^4}{3} \left[
             \widetilde{I}_{jk}(\omega_n)\widetilde{I}_{jk}^*(\omega_n) 
	     -\left|\text{tr}\,\widetilde{I}(\omega_n)\right|^{2}
         \right]
         + {\cal O}\left(m^4\right) \, ,\label{eqn:Lm}
\end{eqnarray}
where 
\begin{eqnarray}
    L_{\text{GR}} & \equiv &
    \sum_{n=1}^\infty 
    \omega_n^6 \left[
             \frac{2}{5}\widetilde{I}_{jk}(\omega_n)
	         \widetilde{I}_{jk}^*(\omega_n) 
	     -\frac{2}{15}\left|\text{tr}\,\widetilde{I}(\omega_n)\right|^{2}
         \right] 
\end{eqnarray}
\end{mathletters}
is the usual general-relativistic expression for the radiated power,  
$\text{tr}\,\widetilde{I}$ is the trace of $\widetilde{I}_{jk}$, and
we sum over repeated indices.  
The quantity in the summation of (\ref{eqn:Lm}) is the first correction to 
the general-relativistic expression for the radiated power due to a 
small nonzero graviton mass.  Comparison of
this correction to the observed orbital decay in binary pulsars 
PSR~B1913+16 and PSR~B1534+12 will provide us with a bound on $m$.

\section{Binary Pulsars}
\label{binarypulsars}

The formula (\ref{eqn:Eloss}) for the energy-loss rate of a
gravitational-wave source when the graviton is massive is easily
applied to the orbital decay of binary systems.  Consider two bodies
of masses $M_1$ and $M_2$, orbiting in the $xy$ plane with coordinates
$(d_1\cos(\theta),d_1\sin(\theta))$,
$(-d_2\cos(\theta),-d_2\sin(\theta))$.  Choosing the origin to be at
the center of mass, one has
\begin{mathletters}
\begin{eqnarray}
d_1 &=& \frac{\mu d}{M_1}  \, , \\
d_2 &=& \frac{\mu d}{M_2}  \, ,
\end{eqnarray}
where $d$ is the orbital separation of the binary components, 
$\mu$ is the system's reduced mass, and $M$ is its total mass,
\begin{eqnarray}
    d   & \equiv & d_{1}+d_{2},\\
    \mu & \equiv & {M_{1}M_{2}\over M},\\
    M   & \equiv & M_{1}+M_{2}.
\end{eqnarray}
\end{mathletters}
Assuming a Keplerian orbit, the motion is described by 
\begin{eqnarray}
d &=& \frac{a(1-e^2)}{1+e\cos(\theta)} \, , \\
\frac{d\theta}{dt} &=& {[Ma(1-e^2)]^{1/2}\over d^2} \, ,  
\end{eqnarray}
where $a$ is the semi-major axis and $e$ is the eccentricity of the orbit.
The nonzero quadrupole moments of this system are 
\begin{eqnarray}
I_{xx} &=& \mu d^2 \cos^2(\theta) \, , \nonumber \\
I_{xy} &=& I_{yx} = \mu d^2 \cos(\theta)\sin(\theta)  \, , \nonumber \\
I_{yy} &=& \mu d^2 \sin^2(\theta) \, . 
\end{eqnarray}
The Fourier transform of the quadrupole moment of Keplerian orbits is 
known \cite{peters63a}.  For $n>0$ 
\begin{eqnarray}
\widetilde{I}_{xx}(\omega_{n})
  & = &  \frac{\mu a^2}{2n}\left[
             J_{n-2}(ne) -2eJ_{n-1}(ne) +2eJ_{n+1}(ne)
             -J_{n+2}(ne)
         \right] \, , \label{FTIxx} \nonumber \\
\widetilde{I}_{xy}(\omega_{n}) 
  & = &  i\frac{\mu a^2}{2n}(1-e^2)^{\frac{1}{2}}\left[
             J_{n-2}(ne) -2J_{n}(ne)+J_{n+2}(ne)
         \right] \, , \label{FTIxy} \nonumber \\
\widetilde{I}_{yy}(\omega_{n}) 
  & = &  -\frac{\mu a^2}{2n}[
             J_{n-2}(ne) -2eJ_{n-1}(ne)+\frac{4}{n}J_{n}(ne) 
	     +2eJ_{n+1}(ne)-J_{n+2}(ne)
         ] \, , \label{FTIyy} 
\end{eqnarray}
where the $J_{n}(x)$ are Bessel functions of the first kind
The moments for $n<0$ follow from 
\begin{equation}
    \widetilde{I}_{jk}(\omega_{-n})= 
\widetilde{I}_{jk}^*(\omega_{n}).
\end{equation}

Combining these quadrupole moments with equation (\ref{eqn:Eloss})
provides us with an easy means to put a limit on the graviton mass. 
For example, the orbital decay rate of the binary pulsar PSR~B1913+16 has been 
measured and found to be slightly in excess of the predictions of general
relativity \cite{taylor94a}.  
Denote by $P_{\text{b}}$ the measured orbital period of the binary system, 
$\dot{P}_{\text{b}}$ the measured orbital period derivative ascribed to 
gravitational radiation, and $\dot{P}_{\text{GR}}$ the instantaneous 
period derivative expected owing to general-relativistic 
(i.e., zero graviton rest mass) orbital decay.
Identify the fractional discrepancy between the observed 
and predicted decay rates: 
\begin{equation}\label{deltaP}
\Delta 
 \equiv \frac{\dot{P}_{\text{b}}-\dot{P}_{\text{GR}}}{\dot{P}_{\text{GR}}} \, .
\end{equation}
For a slowly decaying Keplerian binary, the instantaneous period derivative 
is proportional to the energy-loss rate; hence, 
\begin{equation}\label{fracdisc}
\frac{\dot{P}_{\text{b}}-\dot{P}_{\text{GR}}}{\dot{P}_{\text{GR}}} 
  =  \frac{L - L_{\text{GR}}}{L_{\text{GR}}} \, ,
\end{equation}
where $L$ is the gravitational-wave luminosity inferred 
from $\dot{P}_{\text{b}}$, 
and $L_{\text{GR}}$ is the energy-loss rate expected from general relativity.  
This quantity has been measured for PSR~B1913+16 and PSR~B1534+12 
(see \cite{taylor94a,stairs99a} and Table~\ref{tbl:pulsar}).

Now suppose that $\Delta$ is due at least in part to a
nonvanishing graviton mass (rather than simply experimental
uncertainties).  Combining (\ref{eqn:Eloss}) and (\ref{fracdisc}), 
this implies an upper limit to the squared graviton mass of
\begin{equation}\label{bound}
m^{2} 
  \le  \frac{24}{5}\,F(e)\,
       \left(\frac{2\pi\hbar}{c^{2}P_{\text{b}}}\right)^{2} \, 
       \frac{\dot{P}_{\text{b}}-\dot{P}_{\text{GR}}}{\dot{P}_{\text{GR}}} \, ,
       \label{eqn:dl/l}
\end{equation}
where $F(e)$ is a function of the eccentricity,
\begin{equation}
F(e) 
  =  \frac{1}{12}\frac{
     \displaystyle{\sum_{n=1}^\infty} \,n^6 \left[
         3\widetilde{I}_{jk}(\omega_n)\widetilde{I}_{jk}^*(\omega_n) 
         -\left|\text{tr}\,\widetilde{I}(\omega_n)\right|^{2}
     \right]
     }{
     \displaystyle{\sum_{n=1}^\infty} \,n^4 \left[
         \widetilde{I}_{jk}(\omega_n)\widetilde{I}_{jk}^*(\omega_n)
	 -\left|\text{tr}\,\widetilde{I}(\omega_n)\right|^{2} 
     \right]
     }  \, .
     \label{eq:F(e)}
\end{equation}
These sums can be performed using the techniques of \cite{peters63a},
giving
\begin{equation} \label{eq:F(e)summed}
F(e) 
 =  \frac{1+\frac{73}{24}e^2+\frac{37}{96}e^4}{(1-e^2)^3}\, .
\end{equation}

\begin{figure}
    \epsfxsize=\columnwidth
    \epsfxsize=5in
    \epsffile{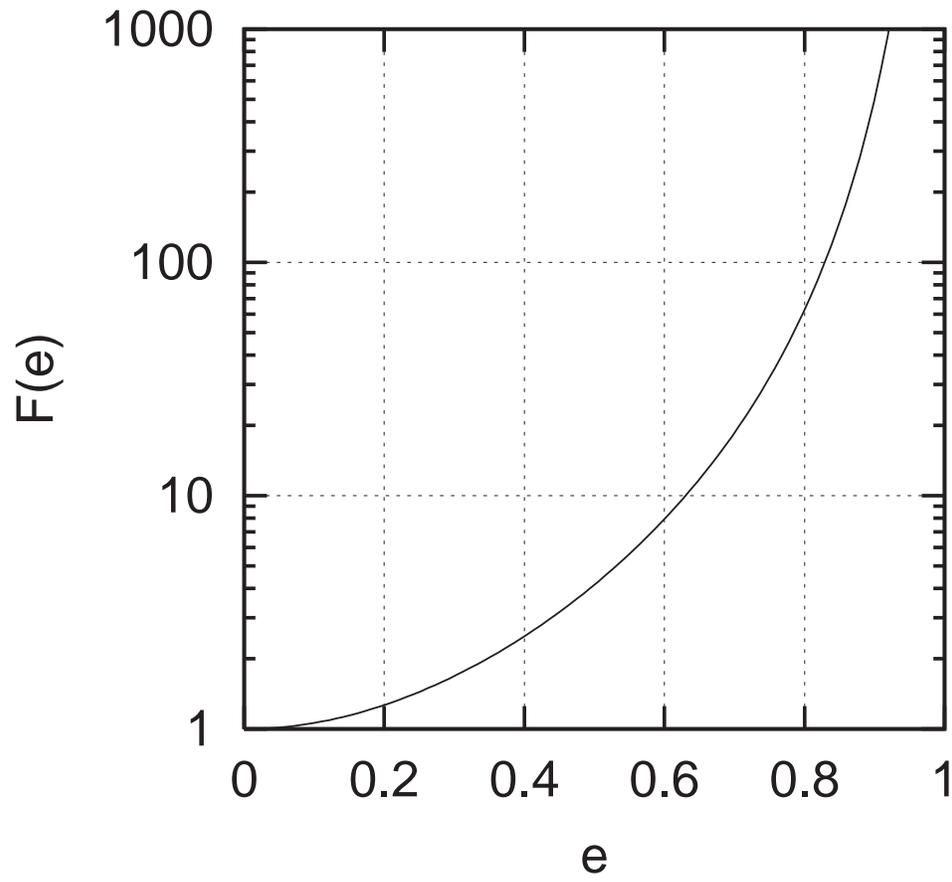}
\caption{Eccentricity factor $F(e)$ (cf.\ eqn.\ \ref{eq:F(e)summed}) versus $e$.}
\label{fig:F}
\end{figure}

The function $F(e)$ is plotted in Figure~\ref{fig:F}.  Note that
$F(e)$ is greater than or equal to unity; a nonzero graviton mass
increases the energy emission of Keplerian binaries, as one would
expect from adding extra degrees of freedom to the gravitational
field.  Figure \ref{fig:F} contains another lesson, as well.  Note
that, for binaries of fixed period, stronger bounds arise from
binaries with smaller eccentricity.  This dependence is easily
understood.  Binaries with large eccentricities have strong speed
variations, as they move from periastron to apiastron.  These speed
variations lead high-eccentricity binaries to produce the bulk of
their radiation in ever higher harmonics of the orbital frequency
\cite{peters63a}.  The effects of a non-zero graviton mass are more
pronounced for lower-frequency gravitational waves, as in equation
(\ref{speed}).  As a result, the ideal system for bounding the graviton
mass is a binary with a large orbital period and small eccentricity (a
weak emitter of gravitational waves), but which still has a measurable
inspiral rate.

Equation (\ref{eqn:dl/l}), relating the squared graviton mass to the 
fractional discrepancy $\Delta$ in the period derivative 
[or equivalently by (\ref{fracdisc}) the fractional discrepancy in the 
luminosity], assumes that this discrepancy is known exactly.  
In fact, the period-derivative discrepancy is
known only up to the errors associated with the measured changes in
the binary period and the acceleration of the binary relative to
Earth.  In practice, the one-sigma uncertainty in $\Delta$ (which 
is listed in Table~\ref{tbl:pulsar}) is of the same order as the
measured discrepancy and must be accounted for, as the actual $\Delta$ 
could reasonably be expected to differ from the value derived
from the measurements by one or more standard deviations. 
Consequently, we must describe the actual upper-limit on the mass
statistically.  In the absence of detailed information we assume the
measured discrepancy $\Delta$ to be normally distributed about its unknown 
actual value [given by the equality in (\ref{bound}) with unknown $m^2$], 
and with standard deviation as given in Table \ref{tbl:pulsar}.  In our
model we relate the discrepancy to the squared graviton mass, which
must be non-negative.  Referring to \cite[Table X]{feldman98a}, which
lists the 90\% unified upper limit/confidence intervals for the
non-negative mean of a univariate normal distribution 
based on a measured sample from the distribution, we calculate the
90\% upper limit on the (non-negative) graviton mass, which is given
in the final row of Table \ref{tbl:pulsar}.

The best single limit on the graviton mass,
$m<6.4\times10^{-20}\,\text{eV}/c^2$, comes from the observations of
PSR~B1534+12.  This is despite the larger uncertainty in the measured
luminosity discrepancy, compared to PSR~B1913+16, because the
luminosity discrepancy for PSR~B1534+12 is negative.  A negative
discrepancy, taken as exact, would correspond to a negative graviton
mass, which is unphysical.  Correspondingly, a negative measured
discrepancy is particularly unlikely to arise from a positive $m^{2}$
compared to a vanishing $m^{2}$, which leads to a tighter upper limit.

We may combine the two observed discrepancies to find a single upper 
bound on the graviton mass. Each observation $k$ results in a 
discrepancy $\Delta_{k}$ and an associated one-sigma uncertainty 
in the estimated discrepancy $\sigma_{\Delta,k}$. These in turn are
related, through equation (\ref{eqn:dl/l}), with measurements of $m^{2}_{k}$, 
together with associated one-sigma uncertainties $\sigma_{k}$. The 
quantity
\begin{mathletters}
\begin{equation}
    m^{2} \equiv {m_{1}^{2}+\beta m_{2}^{2}\over1+\beta}
\end{equation}
is then a normally distributed random variable whose mean is the 
squared graviton mass and whose variance is 
\begin{equation}
    \sigma^{2} = {\sigma_{1}^{2}+\beta^{2}\sigma_{2}^{2}\over
    \left(1+\beta\right)^{2}} \, . 
\end{equation}
Choosing 
\begin{equation}
    \beta = \left({\sigma_{1}\over\sigma_{2}}\right)^2  
\end{equation}
minimizes the variance of $m^{2}$: 
\begin{equation}
    \sigma^{2} = {\sigma_{1}^{2}\sigma_{2}^{2}\over
    \sigma_{1}^{2}+\sigma_{2}^{2}} \, .
\end{equation}
\end{mathletters}
Referring to Table \ref{tbl:pulsar} and \cite[Table X]{feldman98a}, 
the corresponding limit on the graviton mass from the combined 
observations of PSR~B1913+16 and PSR~B1534+12 is thus
\begin{equation}
    m_{90\%} < 7.6\times10^{-20}\,\text{eV}/c^2.
\end{equation}

\section{Discussion}
\label{conclusions}

Table \ref{tbl:pulsar} gives the relevant parameters and the
corresponding graviton mass bounds for the two binary pulsars whose
gravitational-wave induced orbital decay has been measured, PSR~B1913+16 
and PSR~B1534+12 \cite{taylor94a,stairs99a}.  The graviton
mass bounds from the timing observations of each system are very
similar, and about two orders of magnitude weaker than the Yukawa
limit obtained from solar-system observations, $m< 4.4 \times
10^{-22}\,\text{eV}/c^2$ \cite{talmadge88a}.  Both of these bounds are, in turn,
several orders of magnitude weaker than that provided by observations
of galactic clusters, $m<2\times10^{-29}\,\text{eV}/c^2$
\cite{goldhaber74a,hare73a}, though we regard these galactic cluster
bounds as less robust, owing to their reliance on assumptions about
the dark matter content of the clusters, for example.  In contrast,
the bound obtained here is very straightforward and involves few
assumptions, making it less prone to error: the chief assumption that
we have made is the form of the effective mass term for the graviton,
which --- while not unique --- is natural.  Furthermore, any other
mass term would be expected from dimensional arguments to yield
similar results.

We have assumed that only measurement errors enter into the
determination of the intrinsic binary period decay rate $\dot{P}_{b}$. 
In fact, the determination of this rate requires an estimate of the
acceleration of the binary system, which is principally toward the
galactic center \cite{stairs99a}.  This, in turn, depends on an 
accurate distance measurement to the binary system, which can be
difficult to make.  A systematic error in this distance estimate leads
directly to an error in the estimated acceleration of the binary and,
in turn, to an error in the $\dot{P}_{b}$ ascribed to gravitational
radiation induced decay of the binary system.  The large uncertainty
in the discrepancy $\Delta$ associated with PSR~B1534+12 may well be
due to an underestimate of the distance to this binary system
\cite{stairs99a}, in which case the bound on $m^{2}$ would be even
tighter.

The bound described here arises from the properties of dynamical
relativity, making it conceptually independent of either the solar
system or galactic cluster bounds on the graviton mass, which
are based on the Yukawa form of the static field in a massive theory. 
Furthermore, we expect improvement in the bounds from any given pulsar
system as observations improve the accuracy of the measured fractional 
discrepancy in the period derivative.  For example, when the observations 
of PSR~B1534+12 improve limits on $\Delta$ to the same level as is 
observed today for PSR~B1913+16, the corresponding single-system bound 
on the graviton mass should improve to 
approximately $2\times10^{-20}\,\text{eV}/c^2$.

The field of gravitational-wave detection is new.  We are only just
now learning to exploit the opportunities it is creating for us. 
Within the next year, several large ground-based interferometric
detectors will begin full operation \cite{luck00a,coles00a,marion00a},
and existing cryogenic acoustic detectors
\cite{hamilton97a,cerdonio97a,astone97a,blair00a} will see significant
improvements in sensitivity.  Within the next decade we should see
further enhancements in the capability of all these instruments
\cite{M990288A1,dewaard00a,cerdonio00a}, and the deployment of the
space-based interferometric detector LISA (Laser Interferometer Space
Antenna) \cite{lisa1,lisa2}.  As gravitational-wave observations
mature, we can expect more and greater recognition of their utility as
probes of the character of relativistic gravity.  The opening of the
new frontier of gravitational-wave phenomenology promises to be an
exciting and revealing one for the physics of gravity.

 
\section*{Acknowledgments}

The authors are grateful to Valeri Frolov, Matt Visser, Joel Weisberg, 
Cliff Will, Alex Wolszczan, and Andrei Zelnikov for helpful discussions.  
PJS would like to thank the Natural Sciences and Engineering Research 
Council of Canada for its financial support.  This work has been funded 
by NSF grant PHY~00-99559 and its predecessor.  The Center for Gravitational
Wave Physics is supported by the NSF under co-operative agreement
PHY~01-14375.

  
\appendix 
\section{Choice of Mass Term}
\label{massterms}

The most general mass term possible for the linearized action
(\ref{action}) is proportional to
$\left[h_{\mu\nu}h^{\mu\nu}-\kappa\,(h^\nu_{\hphantom{\nu}\nu})^2\right]$,
with $\kappa$ an arbitrary constant.  Here we demonstrate that
$\kappa=\frac{1}{2}$ is the unique choice possessing both of the
following properties:
\begin{enumerate}
    
    \item the field equations for the metric perturbations can be
    written in the standard form
    \begin{equation}\label{desiredform}
	(\Box-m^2)h_{\mu\nu} = -16\pi\,T^{\mbox{\tiny{eff}}}_{\mu\nu} \, , 
    \end{equation}
    where the source $T^{\mbox{\tiny{eff}}}_{\mu\nu}$ is a local
    function of the stress tensor and is independent of $h_{\mu\nu}$;
    and 
    
    \item taking the limit $m\to0$ in the massive theory recovers the
    predictions of general relativity.
\end{enumerate}
The first property is practical, while the second is necessary for
agreement with experiment.

The field equation for $h_{\mu\nu}$ for general $\kappa$ is   
\begin{equation}
\Box h_{\mu\nu} 
- h_{\mu\hphantom{\lambda},\lambda\nu}^{\hphantom{\mu}\lambda}  
- h_{\nu\hphantom{\lambda},\lambda\mu}^{\hphantom{\nu}\lambda} 
+ h_{,\mu\nu} 
+ \eta_{\mu\nu} h^{\rho\sigma}_{\hphantom{\rho\sigma},\rho\sigma}
- \eta_{\mu\nu} \Box h
- m^2(h_{\mu\nu}-\kappa\,\eta_{\mu\nu} h)
  =  -16\pi\,T_{\mu\nu}  \, . \label{genmassFE}
\end{equation}
The divergence of both sides of (\ref{genmassFE}) must be equal, implying  
\begin{equation}\label{divergencecondition}
h^{\mu\nu}{}_{,\nu} = \kappa\,h^{,\mu}  \, . 
\end{equation}
Taking the trace of the field equation and using this divergence 
condition gives the trace condition 
\begin{equation}\label{tracecondition}
2(1-\kappa)\Box h + (1-4\kappa)m^2 h = 16\pi\,T^{\nu}_{\hphantom{\nu}\nu} \, . 
\end{equation}
We see that $h$ can only be written as a local function of the stress 
tensor if $\kappa=1$. 

Substituting the trace and divergence conditions into the field equation gives
\begin{equation}
(\Box-m^2)h^{\mu\nu} 
  =  -16\pi\left(
         T^{\mu\nu}
	 -\frac{1}{2}\,\eta^{\mu\nu}T_{\lambda}^{\hphantom{\lambda}\lambda}
     \right) 
     + (2\kappa-1)\left[
         h^{,\mu\nu} 
         +\frac{1}{2}\,\eta^{\mu\nu}m^2 h
     \right] \, , 
\end{equation}
which is of the desired form (\ref{desiredform}) except for the 
term in square brackets. The latter can be removed only for two 
special values of $\kappa$.   
For $\kappa=\frac{1}{2}$ the coefficient vanishes, leaving 
\begin{equation}
(\Box-m^2)h^{\mu\nu} 
  =  -16\pi\left(
         T^{\mu\nu}
	 -\frac{1}{2}\,\eta^{\mu\nu}T_{\lambda}^{\hphantom{\nu}\lambda}
     \right) 
     \, , 
\end{equation}
which is equivalent to (\ref{fieldeqn}).  
For $\kappa=1$ (the Pauli-Fierz mass term used by Boulware 
and Deser \cite{boulware72a}) we can use the trace condition 
(\ref{tracecondition}) to rewrite the term in square brackets  
as a local function of the stress tensor, yielding  
\begin{eqnarray}
(\Box-m^2)h^{\mu\nu} 
    =    -16\pi\left(
             T^{\mu\nu}
	     -\frac{1}{3}\,\eta^{\mu\nu}T_{\lambda}^{\hphantom{\nu}\lambda}
             +\frac{1}{3m^2}\,T_{\lambda}^{\hphantom{\lambda}\lambda,\mu\nu}
         \right) \, , \label{PauliFierz}
\end{eqnarray}
which is also of the desired form (\ref{desiredform}).  
It is well known, however, that the predictions of 
the $\kappa=1$ theory do not reduce to those of general relativity 
for $m\to0$: this is the van Dam-Veltman-Zakharov discontinuity 
\cite{VaVe:70,Za:70}.  
We are thus led to the choice $\kappa=\frac{1}{2}$ 
and the massive graviton theory described by (\ref{action}).


\begin{table}
  \caption{Orbital parameters and corresponding graviton mass bound 
  from the two binary pulsar systems whose gravitational wave 
  induced orbital decay has been measured. Pulsar parameters are 
  taken from \protect\cite{taylor94a,stairs99a}; see also \protect\cite{Ka:01}. 
  One-sigma uncertainties are quoted for $\Delta$.}
  \label{tbl:pulsar}  
  \begin{tabular}{l|rr}
    & PSR~B1913+16           & PSR~B1534+12 \\
    \hline
    Period & 27907~s                & 36352~s \\
    Eccentricity & 0.61713                & 0.27368 \\
    $\Delta$ & 0.32\% $\pm$ 0.35\%    & $-12.0$\% $\pm$ 7.8\% \\
    Graviton mass 90\% upper bound& $9.5\times10^{-20}\,\text{eV}/c^2$ & 
    $6.4\times10^{-20}\,\text{eV}/c^2$ \\
  \end{tabular}
\end{table}

\end{document}